\documentclass{iopjournal}
\usepackage{graphicx}%
\usepackage{amsmath,amssymb,amsfonts}%
\usepackage{amsthm}%
\usepackage{mathrsfs}%
\usepackage[caption=false]{subfig}
\usepackage{booktabs}
\usepackage{cite}
\usepackage{hyperref}

\usepackage{xurl}

\begin{document}

\articletype{Paper} 

\title{Understanding entropy production via a thermal zero-player game}

\author{M S\"uzen $^{1,2}$ \orcid{0000-0002-9460-7297}}

\affil{$^1$Member, American Physical Society, College Park, Maryland, United States} \\
\affil{$^2$Resident Scientist, Assia, CY 5561, Cyprus}

\email{mehmet.suzen@physics.org}

\keywords{entropy production, stochastic thermodynamics, 
lattice gases, stochastic games}

\begin{abstract}
Understanding the natural bounds of entropy production for driven 
nonequilibrium dynamics in many-body systems reveals how the 
fundamentals of thermodynamics manifest in these regimes across 
a wide variety of systems. In this direction, we propose and study 
the dynamics of a thermal zero-player entropy game, the Ising–Conway 
Entropy Game (ICEg), a self-driven system exhibiting characteristics 
of lattice gases, Ising models, and discrete games. We show that there 
is a universal bound on the entropy production rate, independent of 
temperature and lattice size. The thermalized game is shown to be 
physically interesting and a plausible testbed for studying the 
fundamentals of stochastic thermodynamics.
\end{abstract}

\section{Introduction}  \label{intro}

The second law of thermodynamics dictates that systems evolve toward 
states of maximal entropy under given constraints, tending from ordered 
to disordered configurations \cite{gibbs, kittel04, callen}. In nonequilibrium 
regimes, this principle manifests through the rate of entropy production, 
which becomes the central quantity governing the dynamics. Here, the minimum 
entropy production principle, which holds in certain linear response 
regimes, provides a useful framework for understanding non-equilibrium 
steady states. In stochastic thermodynamics \cite{dorfman, peliti21}
as it is now a central theme in generative artificial intelligence \cite{welling26}, 
the definition and measurement of entropy production remain an active 
area of research \cite{maes00, tiet06z, landi21, liu25, bros26}.

Example definitions of entropy production (EP), such as the correlation 
between system and environment \cite{esposito10}, its connection to free 
energy differences \cite{jarzynski97}, and the recent variance sum 
rule \cite{terlizzi24}, are among the most prominent. Current interest 
in EP has emerged in active matter \cite{agranov22, bebon25, huang26, knight26}, 
nonequilibrium magnets \cite{tie25}, optical matter \cite{chen24}, interacting particle 
systems \cite{bros26}, quantum dots \cite{shen26}, microwave cavities \cite{edet25}
and within fluctuation theorems \cite{wu26}. Typical 
testbeds for studying EP often involve complex dynamics, such as those 
governed by the Fokker–Planck–Kramers equation \cite{fio19}. In this 
context, we propose a physically accessible model: the thermal 
Ising–Conway Entropy Game (ICEg) \cite{suzen23}. We further introduce 
a tractable definition of EP and entropy in lattice dynamics, based on 
the ratio of entropy increase to low-temperature energy, integrated over 
time. The concept of entropy increase and its measurement in lattice 
systems has roots in early work by Linus Pauling \cite{pauling35}, and 
was later formalized in stochastic thermodynamics through Anderson’s 
seminal contributions \cite{anderson58, lagendijk09}. The literature 
highlight the pedagogical value of lattice systems in elucidating entropy 
increase—from chemical bonding to fundamental physical 
principles \cite{susskind, aastrand24}.

Beyond studies of entropy production (EP), the ICEg, as a zero-player game, 
offers a system for exploring game-theoretic approaches to stochastic 
thermodynamics, merging features of lattice gas (cellular automaton) 
models and spin-glass systems. As its name indicates, local energies are 
defined in an Ising-model form on lattice sites \cite{ising}, and the 
evolution rules follow a cellular automaton scheme reminiscent of Conway’s 
Game of Life \cite{gardner70a, gardner70b, gardner71a, gardner71b, 
gardner71c, gardner72a, wolfram83}. While thermalization and fluctuation 
properties of cellular automata inspired by Conway’s rules have been 
studied previously \cite{nina98, adac04, schulman78a}, the thermal 
ICEg provides a simplified yet physically meaningful framework for 
investigating fluctuation dynamics in discrete, lattice-based systems.

The origins of lattice gases date back to seminal work by Lee and 
Yang \cite{leeyang52b}, who first noted the similarity between lattice 
gases and the Ising spin model. The occupation or vacancy of lattice 
sites corresponds to spin orientations up or down. In this context, the 
statistical mechanics of a one-dimensional lattice gas was 
established \cite{ruelle68}. Similarly, lattice gas models in fluid 
dynamics were introduced \cite{frisch86, dhu86} and later generalized 
to the Lattice Boltzmann method \cite{mc88, he97}. In this perspective, 
ICEg contributes to the lattice gas community as a testbed system.

ICEg dynamics extend the occupation of lattice gases into a game-theoretic 
perspective. While game theory \cite{von47} and lattice systems have been 
extensively studied for cooperation dynamics \cite{sza98, sza16, luc22}, 
the movement of occupied sites—resembling hopping in the model—mirrors 
similar mechanisms in ICEg. This feature classifies ICEg as a zero-player 
game and positions it as an ideal tool for studying thermodynamics and 
game-theoretic dynamics. Given the established similarities between 
statistical mechanics and game-theoretic equilibrium \cite{flem04, hauert05, 
babajanyah20}, as well as from an evolutionary perspective \cite{adami18}, 
phase transitions \cite{scott22, muk25} and entropy production in 
games \cite{kharazmi23, fujimoto24} can be explored within this framework.

On the other hand, introducing a thermal bath in molecular dynamics 
is widely studied in the literature \cite{hunenberger05}; approaches 
typically modify the dynamical equations to appropriately introduce 
temperature. A zero-player game like ICEg, using the Monte 
Carlo \cite{binder2010monte, newman1999monte} approach—where moves 
are accepted based on temperature—is well-suited. Using 
Metropolis \cite{metropolis1953} and Glauber \cite{glauber1963, suezen14, suzen26ps} 
spin-flip dynamics \cite{binder2010monte, newman1999monte} is feasible. 
This approach goes beyond a simple noise term, instead inducing the 
dynamics that samples the given energy landscape \cite{doye99, wales06} 
corresponding to the system’s temperature in thermal equilibrium. 
This novel framework enables the study of a game within a statistical 
physics context \cite{babajanyah20, fujimoto24}.

Our primary finding is that the thermal ICEg game dynamics exhibit a novel 
phenomenon: entropy production is naturally bounded in driven nonequilibrium 
thermodynamic systems. This is supported by finite-size scaling analysis showing 
data collapse, and the universality of the results regardless of temperature and 
the geometric properties of the lattice.

The thermal ICEg game serves as a physically plausible test bed at the intersection
of lattice gases, game theory, and self-driven nonequilibrium dynamics: (1) We 
simplify the analysis of the thermalization process in a realistic setting, where 
physical systems evolve self-driven toward a nonequilibrium state from sufficiently 
ordered initial conditions. For this reason, we adopt a Hamiltonian canonical ensemble 
via a thermal bath scheme. This approach is amenable to numerical simulations, 
enabling the construction of an informal analogy between player utilities—originally 
defined in game theory—and entropy maximization during system evolution, as recently 
studied \cite{babajanyah20}. (2) We introduce a measure for entropy production as 
a function of temperature, without relying on non-equilibrium free energy 
theorems \cite{jarzynski97} within stochastic thermodynamics \cite{peliti21}. 
These contributions enable researchers to conduct numerical experiments as a 
test bed for more sophisticated theoretical analysis, demonstrating a universal 
upper bound on entropy production. The bound arises naturally from the system's 
dynamics and does not impose additional constraints on the evolution, as in standard 
Markov chain models \cite{nishi23}.

In Section \ref{gamerules}, we explain the rules of the lattice game, which are 
simpler than lattice gas models and do not impose constraints on transition rates. 
Importantly, we introduce a thermal bath scheme with a fixed initial condition: 
initially, a corner is occupied, starting from a highly ordered state. 
In Section \ref{gentropy}, we formulate the computation of entropy and its relation 
to lattice dynamics. We then describe the resulting entropic regime transition 
observed in numerical simulations. In Section \ref{eprod}, we examine how entropy 
production depends on game dynamics and temperature. In Section \ref{fss}, we 
demonstrate coupled finite-size scaling, which reveals a universal upper bound 
on the entropy production rate. The final section provides conclusions and outlook.

\section{Thermal Zero Player Game}\label{gamerules}

The Ising–Conway Entropy Game (ICEg) \cite{suzen23} is a new type of system that 
shares features with lattice gases, Ising-type models, and lattice games. 
The dynamics are defined on a one-dimensional configurational setting, where 
lattice site occupancy and zero-player game rules govern the evolution of 
configurations. When coupled to a heat bath, the system represents a model 
in the Gibbsian canonical ensemble. The distinct initial condition—occupancy 
at a corner—drives the system from an ordered state toward nonequilibrium 
statistical mechanics. We detail these aspects here for use in the study of 
entropy production (EP).

\begin{enumerate}
\item {\it Similarity to Lattice Gases}: $N$ sites (cells) are arranged in a one-dimensional lattice, 
with each site occupied or unoccupied (analogous to spin up/down). We represent the configuration 
using a binary vector for mathematical convenience.
\item {\it Initialization}: An initial state is prepared by occupying $M$ corner sites,
i.e., a sequence of $1$s at the end of the lattice. For example, for $N=10$ and $M=4$, 
the configuration may be represented as the binary vector  $111100000$ or $000001111$, depending 
on whether the left or right corner is taken as the reference.
\item {\it Similarity to Spin-flip Dynamics} (Moves):
Analogous to Ising model spin-flip dynamics, a randomly selected site 
is flipped if occupied, in the direction opposite to the initial corner. 
The neighboring site must remain unoccupied. More precisely, the update 
corresponds to hopping to one of the unoccupied neighboring sites. We continue 
to refer to this as a spin-flip because the update involves a single-site 
transition (hopping) per Monte Carlo step. For instance, if site $s_{i}$
is occupied, one of its neighbors is selected at random; if that neighbor 
is unoccupied, the system evolves to the new configuration $s_{i-1}$ or $s_{i+1}$. 
If the selected site is unoccupied, the configuration remains unchanged.
\item {\it Thermalization}: A Monte Carlo procedure is applied to the spin-flip dynamics 
to sample the configuration space at a given temperature.
\item {\it Total Energy}: The total energy of the lattice is computed based on the occupation 
at each site, $N$ sites, indexed from $0$ to $N-1$,
$$H(s_{0}, s_{1},...,s_{N-1})=$$
 \begin{equation}
    \begin{cases}
      \frac{1}{2} { \Large \sum_{i=1}^{N-2} } (s_{i-1}+s_{i+1}), & s_{i}=1, i > 0 \\
      \frac{1}{2} s_{1} & s_{i}=1, i = 0 \\
      \frac{1}{2} s_{N-2} & s_{i}=1, i = N-1 \\
      0, & s_{i}=0  \\
    \end{cases}.
  \end{equation}
Here, $s_{i}$ denotes the occupation state of site $i$. The energy function assigns 
a value to each occupied site based on the states of its next-nearest neighbors. 
The contribution from each occupied site is computed as a sum over its left and 
right neighbors, with each term counted once—effectively, the energy depends 
on the sum $s_{i-1}+s_{i+1}$ for each occupied site at position $i$. This counting 
scheme (from left and right) ensures consistency in the definition of neighbor 
interactions and is essential for capturing per-site dynamics, such as hopping 
to adjacent sites. If a site is unoccupied, its contribution to the total energy 
is zero. The form of the energy function favors configurations in which occupied 
sites are clustered, as such configurations correspond to lower energy. 
\item {\it Metropolis Dynamics}: If a spin-flip is possible at the chosen site, and if one of its neighbors
is unoccupied, we evaluate whether the move is energetically acceptable. Let  $\Delta H(s_{i})$ be the 
energy difference due to the new configuration. The move is accepted if the following condition is satisfied, 
given inverse temperature, $\beta=(k_{B}T)^{-1}$, $\beta > 0.0$ \cite{metropolis1953},
\begin{equation}
min {\large[} 1.0, \exp(-\beta \Delta H(s_{i}) {\large]} > \alpha,
\end{equation}
where $\alpha$ is a uniformly distributed random number generated by the PCG64 generator \cite{pcg}.
\item {\it Glauber Dynamics}: In the case of Glauber dynamics \cite{glauber1963, suezen14},
is given by
\begin{equation}
 min {\large[} 1.0, 1.0/(1+\exp(\beta \Delta H(s_{i})) {\large]} > \alpha.
\end{equation}
\end{enumerate}

\section{Entropy on the lattice}\label{gentropy}

The concept of entropy manifests in diverse contexts, originating in the 
thermodynamics of heat engines \cite{callen}. Shannon entropy \cite{shannon} 
provides a natural analogue to configurational entropy in discrete systems 
and shares formal similarities with Gibbs entropy \cite{gibbs}. In quantum 
mechanics, von Neumann entropy \cite{vonn} quantifies the entropy of quantum 
states in Hilbert space. A further example arises in systems with strong 
gravitational fields: the Bekenstein entropy of a black hole \cite{bekenstein} 
is proportional to its horizon area.

In our context, we take the largest extent of the occupied sites, as determined 
by the lattice dynamics presented above, as a surrogate for the system’s entropy. 
This provides a more physically intuitive analogy than a formal entropy measure 
in nonequilibrium conditions. Although the lattice game is self-driven before 
reaching thermodynamic equilibrium, the configuration defined by the extent of 
occupied sites from one end to the other determines the ensemble space we sample. 
This reflects the fact that the space of occupied sites is not static—it evolves 
with the dynamics. An example of such behavior is a dice where the number of faces 
may change with each throw, depending on thermalized dynamics. In our case, however,
the system is governed by a thermalized game on a one-dimensional lattice.
 
A more concrete and computationally efficient way to identify the extent of occupied sites 
is to compute the maximum and minimum positions of occupied sites using basic binary index 
algebra. For a 1D lattice $L(t)$ of size $N$ with at most $M$ occupied sites ($M < N$), 
we represent the state as a binary vector  $ L(t) \in \{1,0\}^N$, where $1$ indicates 
an occupied site and $0$ an unoccupied one. The measure of entropy $S(t)$  is then defined as:
\begin{equation}
  S(t) = max\mathbb{I}  \left[ L(t) \right] -  min\mathbb{I} \left[ L(t) \right],
\end{equation}
where $\mathbb{I}$ denotes the indicator function mapping the positions of occupied 
sites to their indices. The quantity $S(t)$ corresponds to the difference between the maximum 
and minimum indices of occupied sites, and thus quantifies the spatial extent of
occupied regions.

For example, in the configuration $0101010000$, the occupied sites are at positions 
$\{2,4,6\}$, so $ max\mathbb{I}  \left[ L(t) \right]=6$, $min\mathbb{I} \left[ L(t) \right]=2$,
and $S(t)=4$. This reflects the total number of lattice sites spanned by the occupied region. 
As a measurable and physically intuitive quantity for tracking configurational dynamics in a 
self-driven lattice system, it serves as a surrogate for entropy, a measure of configurational 
disorder.

\begin{figure}[ht!]
\centering
\subfloat[\label{game:glauber}]{\includegraphics[width=0.45\textwidth]{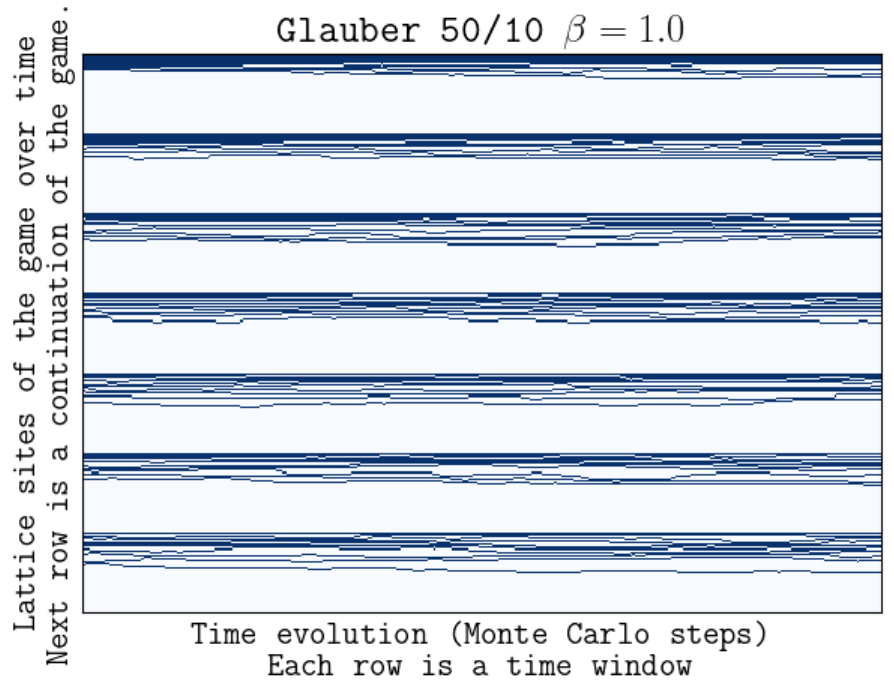}} 
\subfloat[\label{game:metro}]{\includegraphics[width=0.45\textwidth]{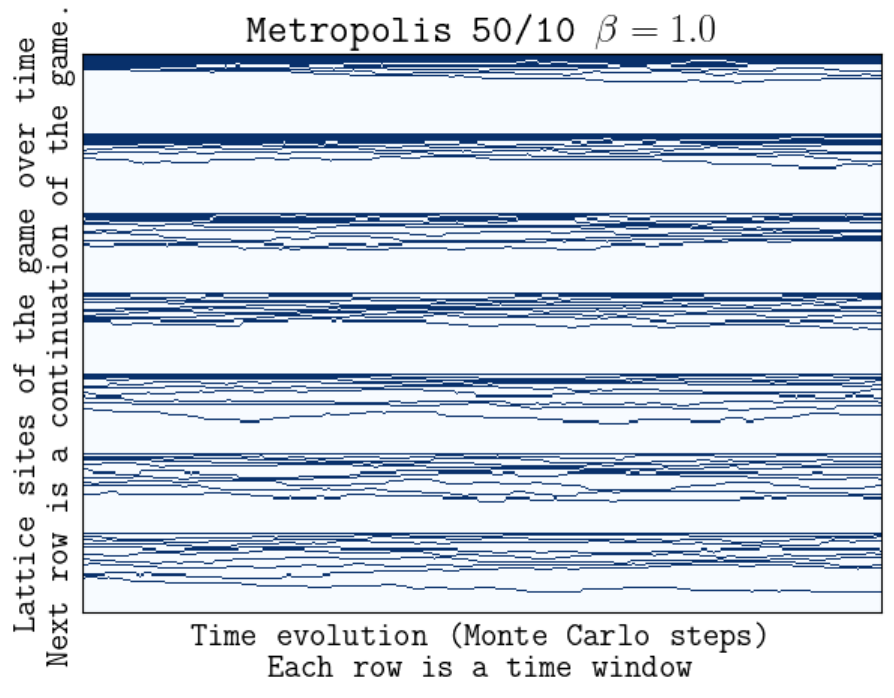}}
\caption{An example evolution of the game is visualized for $N=50$, $M=10$  at $\beta=1.0$.
  Each row represents a time window along the x-axis, with the states of the sites arranged 
  along the y-axis. The next row continues the evolution from the previous row. 
  (a) Glauber Dynamics (b) Metropolis Dynamics}
\end{figure}

We record $S(t)$ at each MC step. We repeat each game play $100$ times with different 
inverse temperatures $\beta=\{0.01, 0.5, 0.9, 1.0, 1.5, 2.0, 5.0, 10.0\}$ and lattice 
settings $(N, M) \in \{ (50, 10), (40,10), (30,10) \}$. The lattice sizes cover different 
proportions of $N$ and $M$. The ratio $N/M$ indirectly reflects the duration for 
which the system remains in non-equilibrium states. In all cases, a clear transition 
to the entropic regime is observed for both Metropolis and Glauber dynamics.

Example evolution of the game for both Metropolis and Glauber dynamics is shown in 
Figure \ref{game:glauber} and Figure \ref{game:metro}, at $\beta=1.0$, $N=50$ and $M=10$. 
The time evolution of lattice sites is presented in a stacked format. We observe how 
diffusion-like behavior emerges over time. Note that, from a stochastic dynamics 
perspective, payoffs correspond to acceptable moves that lead to configurations 
with probabilities governed by Metropolis or Glauber dynamics at the given temperature.

\begin{figure}[ht!]
  \centering
  \subfloat[\label{evol:5010glauber}]{\includegraphics[width=0.45\textwidth]{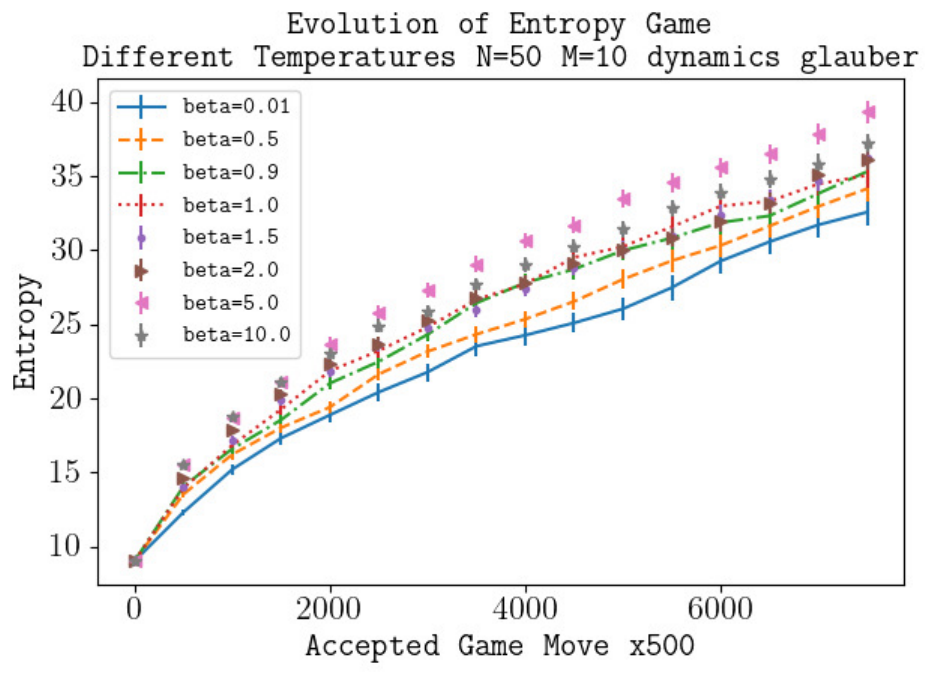}} 
   \subfloat[\label{evol:5010metro}]{\includegraphics[width=0.45\textwidth]{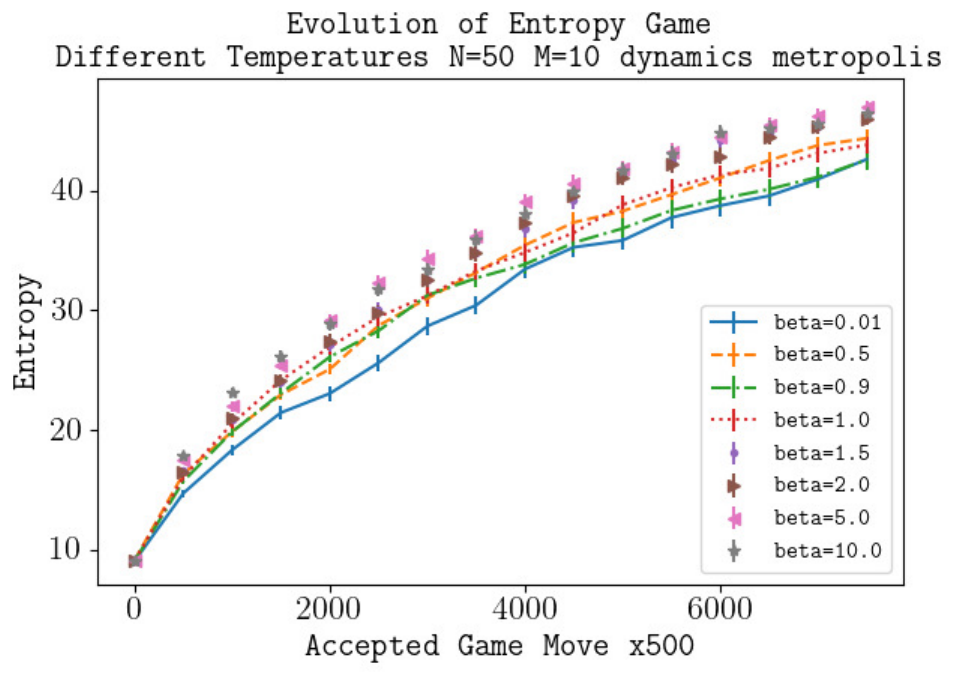}} 
  \caption{
   (a) Evolution of entropy measure for $N=50$, $M=10$ with Glauber dynamics and different inverse temperatures with standard errors.
   (b) Evolution of entropy measure for $N=50$, $M=10$ with Metropolis dynamics and different inverse temperatures with standard errors.
  }
\end{figure}

\begin{figure}[ht!]
  \centering
  \subfloat[\label{evol:4010glauber}]{\includegraphics[width=0.45\textwidth]{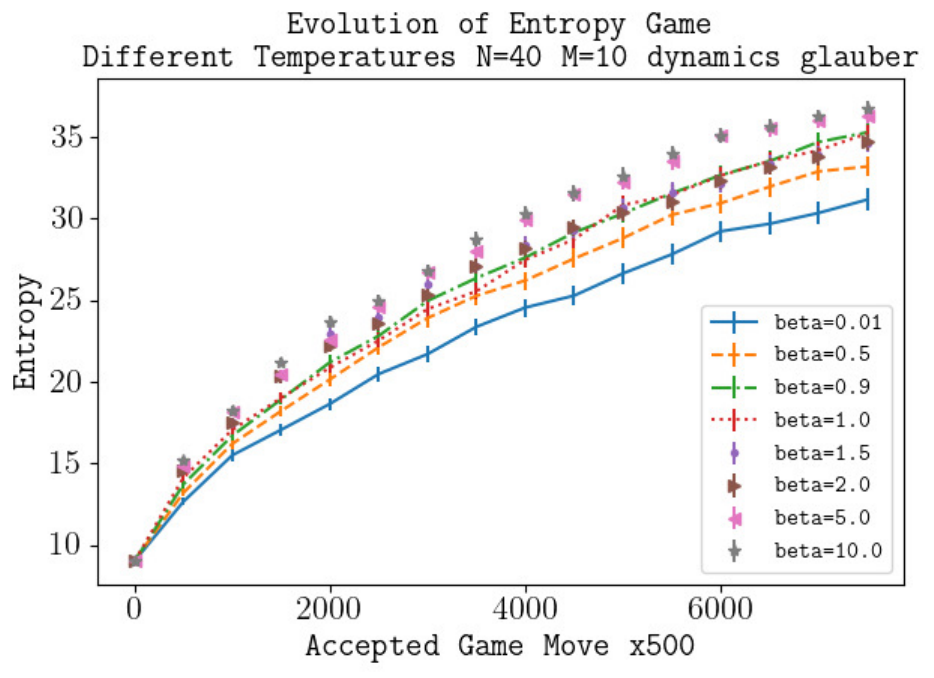}}  
   \subfloat[\label{evol:4010metro}]{\includegraphics[width=0.45\textwidth]{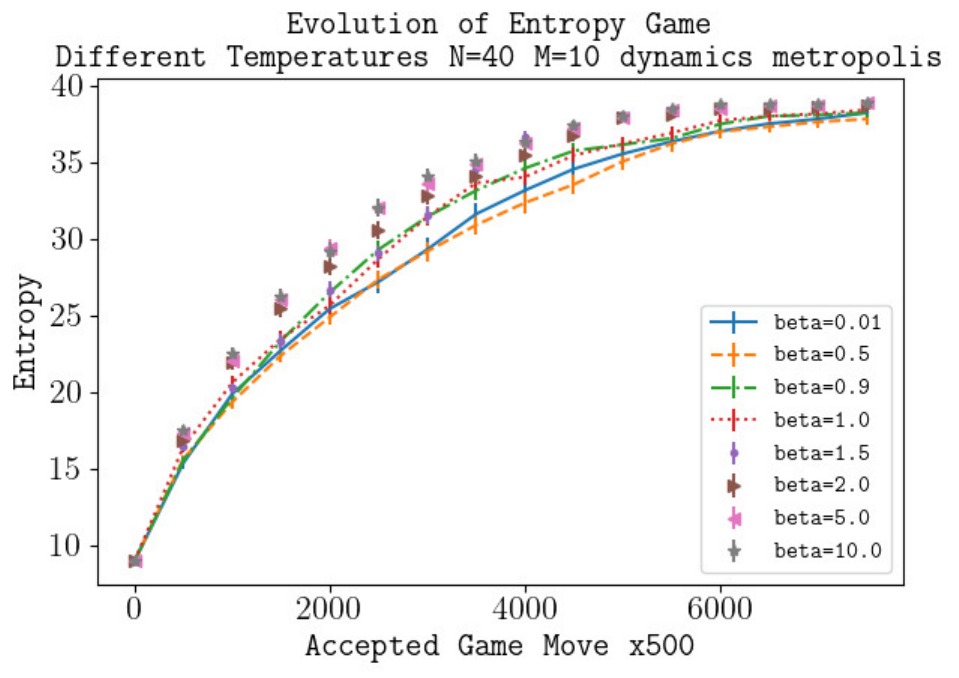}} 
  \caption{
   (a) Evolution of entropy measure for $N=40$, $M=10$ with Glauber dynamics and different inverse temperatures with standard errors.
   (b) Evolution of entropy measure for $N=40$, $M=10$ with Metropolis dynamics and different inverse temperatures with standard errors.
  }
\end{figure}

In Figure \ref{evol:5010glauber} and Figure \ref{evol:5010metro}, the entropy measure over 
time (per game move) is shown for the system size $(N,M)=(50,10$) across a range of temperatures.
The temperature dependence of entropy is more pronounced in the Glauber dynamics: higher 
temperatures lead to greater entropy accumulation, consistent with the energetic expectations 
and validating the simulation results against expected physical behavior. Similar trends are 
observed in Figure \ref{evol:4010glauber} and Figure \ref{evol:4010metro}, where the entropy 
measure over time (per game move) is presented for the system size $(N,M)=(40,10)$, under 
comparable temperature ranges.

\section{Entropy production}\label{eprod}

Entropy production plays a central role in the study of nonequilibrium thermodynamic 
systems. In the case of stochastic lattice games, the concept requires careful 
attention. The definition and interpretation of entropy production vary across 
the literature \cite{prigogine, klein54, jaynes80, maes00, tiet06z, esposito10, landi21}.

We define the entropy production rate as the ratio of the time-integrated entropy production 
over the time-integrated entropy at the minimum temperature:
\begin{equation}
 S_{prod} =  \sum S(t_{0};t;\beta) {\Large(} \sum S(t_{0};t;\beta_{min} {\Large)} )^{-1}.
\end{equation}
This quantity is a normalized proxy for the dissipation in the system, where the denominator 
corresponds to the equilibrium reference case at low temperature. The time-dependent entropy rates 
$S(t_{0};t;\beta)$ are computed along the trajectory of the system's evolution, and the normalization 
ensures that $S_{prod}$ is independent of the baseline temperature.

This is a tractable and intuitive definition, requiring only two distinct computations.
It captures the underlying physical dynamics in nonequilibrium regimes without resorting 
to free energy calculations.

\begin{figure}[ht!]
\centering
  \subfloat[\label{ep:metro}]{\includegraphics[width=0.33\textwidth]{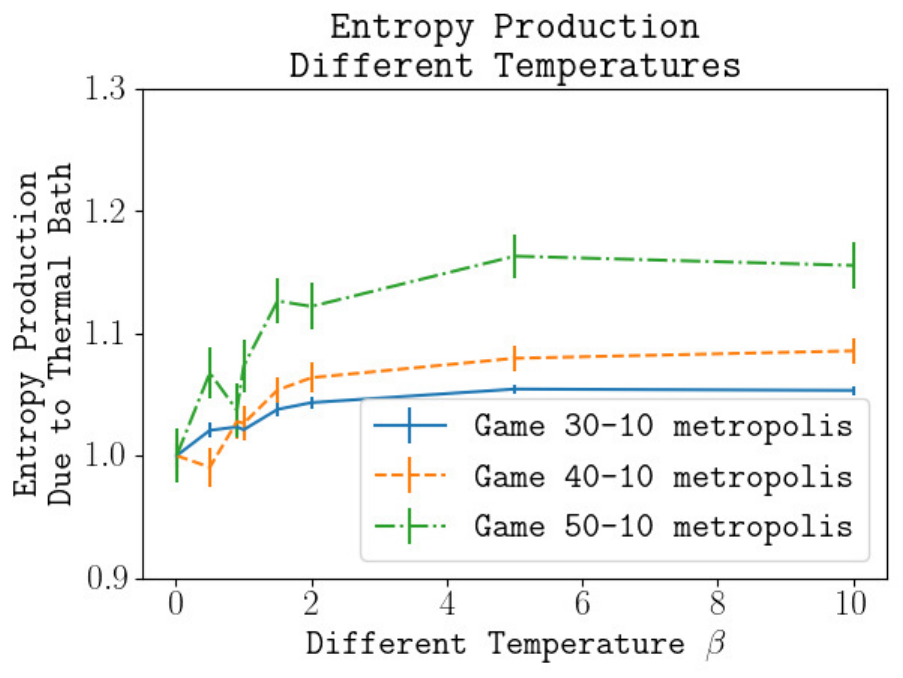}}  
  \subfloat[\label{ep:glauber}]{\includegraphics[width=0.33\textwidth]{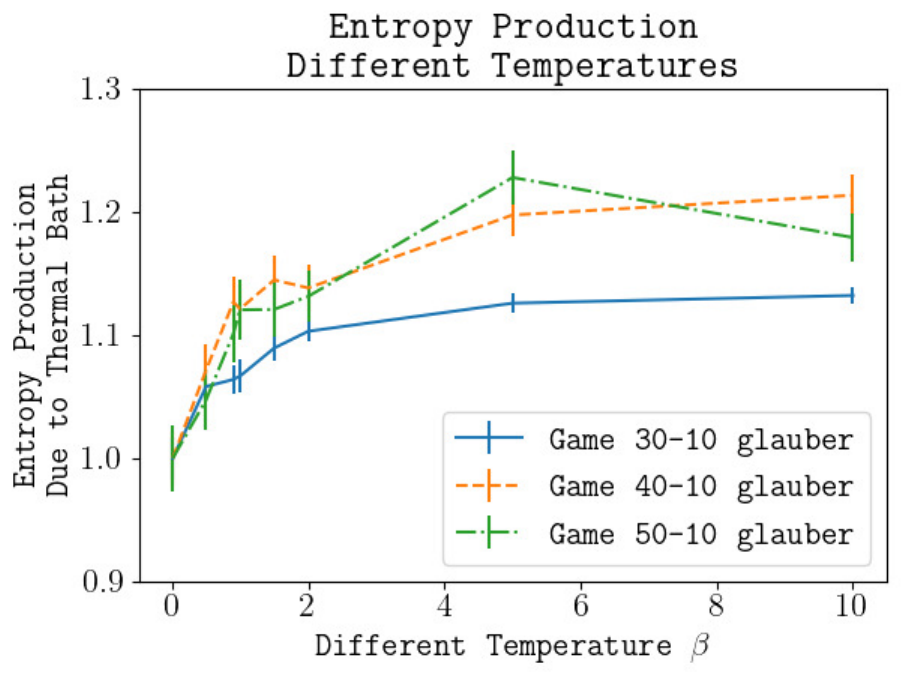}}  
  \subfloat[\label{ep:cdf}]{\includegraphics[width=0.33\textwidth]{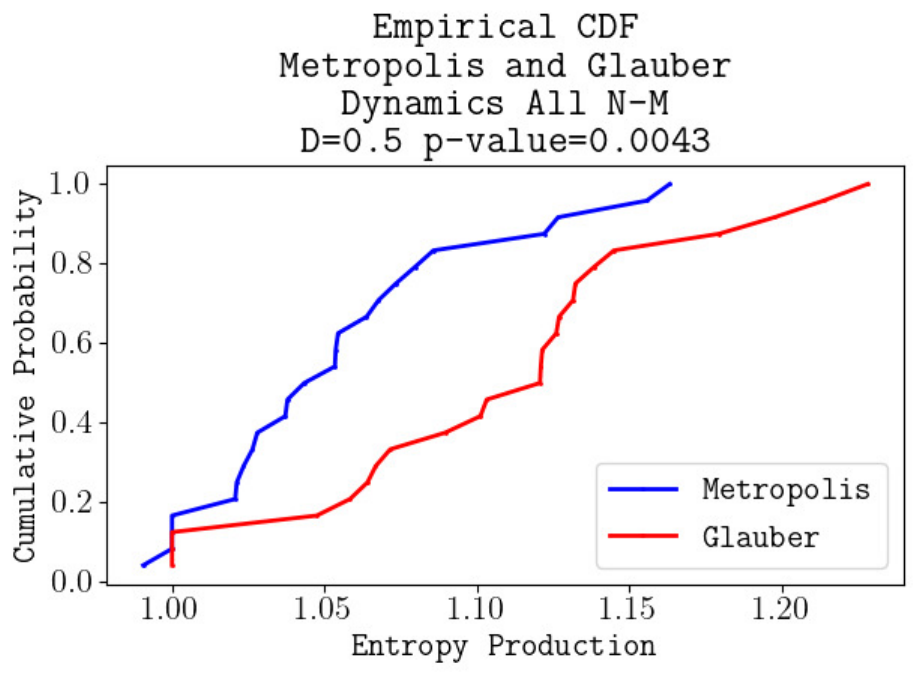}}
  \caption{Entropy production with respect to lowest temperature over range of inverse temperatures and different settings with 
  standard errors. (a) EP for Metropolis dynamics (b) EP for Glauber dynamics (c) Cumulative Distribution 
  Function (CDF) for Metropolis and Glauber EPS, showing statistically significant difference. }
\end{figure}

Entropy production exhibits saturating behavior over temperature ranges for both Metropolis 
and Glauber dynamics, as shown in Figures \ref{ep:metro} and \ref{ep:glauber}. This implies 
that the rate of entropy production in a thermalized system approaches an upper bound as it 
approaches equilibrium. An alternative interpretation is that, regardless of temperature, a 
disordered system eventually settles into a stable state, even if not fully equilibrated, 
due to the dynamics of the underlying update rules. While this behavior may appear 
counterintuitive, it represents a robust demonstration of the second law of thermodynamics: 
in self-driven systems like the Ising–Conway Entropy Game (ICEg), entropy production cannot 
be driven below zero, no matter how large the thermal fluctuations.

From a Monte Carlo sampling perspective, Glauber dynamics is able to capture or resolve 
entropy more effectively than Metropolis dynamics. This may arise because its acceptance 
probability is well bounded. The Kolmogorov–Smirnov (KS) test supports this observation: 
the cumulative distribution functions (CDFs) of entropy production rates and the 
corresponding KS statistics are shown in Figure \ref{ep:metro}

Additionally, the evolution of the game can be summarized as follows: \\
(1) The total number of occupied lattice sites is used as a proxy for the 
system’s entropy. This entropy measure is recorded over both Glauber and 
Metropolis dynamics within the canonical ensemble, across Monte Carlo steps. 
Entropy curves are computed for different temperatures and lattice 
configurations, including the initial number of occupied sites at the 
corner and the extent of free space. These curves are generated across 
a range of temperatures, and their behavior illustrates how entropy 
increases as the system evolves toward a more disordered state. \\
(2) The ratio of the area under these curves to the entropy at the lowest 
temperature represents the total entropy production over the temperature range. 
Separate entropy production curves are computed for different lattice settings 
and for both Glauber and Metropolis dynamics.

\section{Lattice game coupled finite size scaling}\label{fss}

To demonstrate that the observed game dynamics generalize beyond finite system sizes, 
finite-size analysis is required—specifically, finite-size scaling (FSS) \cite{privman}. 
We show that the data collapse (DC) phenomenon \cite{bhattacharjee01} appears in the 
entropy production results across both temperature and lattice size regimes. With the 
growing interest in deep learning training scaling laws \cite{qiu24}, data collapse 
has emerged as a critical diagnostic tool. Its applications have also been reported 
in complex networks \cite{serafino21}, quantum phase transitions \cite{ebadi21}, 
deep diffusion models \cite{biroli24}, quantum information in atomic 
systems \cite{zhang25}, and self-similar dynamics \cite{watan25}.

\begin{table}
\centering
\caption{Fitted exponents and parameters for the scaling form in ICEg dynamics, 
with dynamics are specified by column.}
\begin{tabular}{l|c|c}
\toprule
 & \textbf{Glauber} & \textbf{Metropolis} \\
\midrule
$a$ & 1.2823 & 1.4546 \\
$b$ & -1.7861 & -2.2593 \\
$c$ & -1.0861 & -1.4720 \\
$d$ & 0.0756 & 0.3823 \\
$A$ & -0.0022 & -0.1880 \\
$B$ & 0.9752 & 1.1076 \\
\bottomrule
\end{tabular}
\label{tabfss}
\end{table}

Standard finite-size scaling typically employs a single system size parameter. 
In our case, two independent size parameters are present: the lattice size 
$N$ and the initial occupation size $M$, which corresponds to the number of 
occupied sites. Because occupied sites can diffuse through the empty space 
of length $N-M$, the FSS analysis must account for both $N$ and $M$. This 
reflects a scenario in which the system's dynamics are sensitive to both 
spatial extent and initial configuration, effectively encoding the entropy 
associated with spatial degrees of freedom. This entropy governs the maximum 
duration over which entropy production can persist.

We formulate the finite-size scaling ansatz for lattice size $N$ 
and the initial occupation size $M$, defined as the number of occupied sites 
at the start of the dynamics. We use a fixed initial condition for the 
Ising–Conway Entropy Game (ICEg): all $M$ occupied sites are initially 
located at one end of the lattice. The entropy production scaling ansatz reads:

$$ EP(\beta, N, M) = N^c M^d f(u).$$

$f(u)$ is the scaling function and the scaling variable is $u=\beta N^a M^b$.
The dimensionless scaling exponents $a,b,c$ and $d$ relate
entropy production to systems inherent scales depending on temperature and inverse 
temperature $\beta$. We choose a specific form for the scaling function, which is not 
unique, to be $f(u)=Au+B$, $A$ and $B$ are dimensionless. 

\begin{figure}[ht!]
\centering
  \subfloat[\label{fss:metro}]{\includegraphics[width=0.45\textwidth]{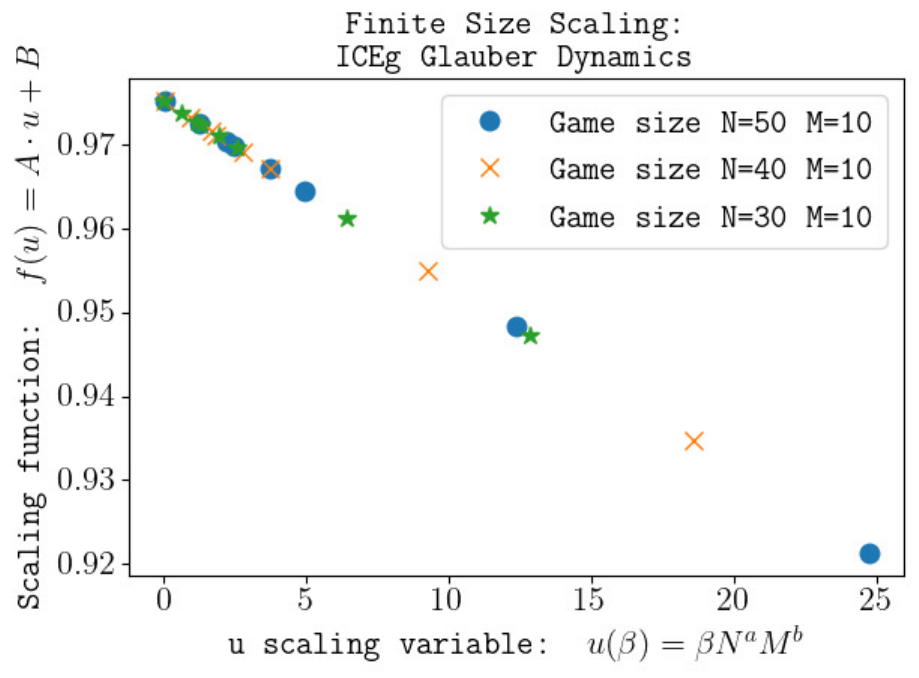}}  
  \subfloat[\label{fss:glauber}]{\includegraphics[width=0.45\textwidth]{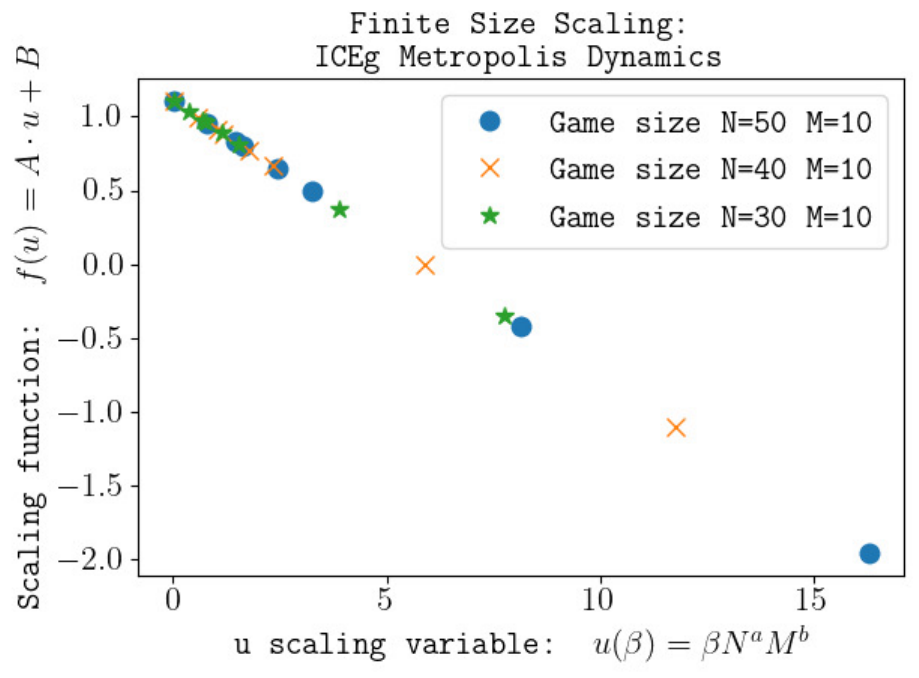}} 
  \caption{Entropy production finite size scaling analysis: 
           (a) Metropolis dynamics (b) Glauber Dynamics}
\end{figure}

We observed data collapse or scaling collapse for both Metropolis and Glauber dynamics 
in Figures \ref{fss:metro} and \ref{fss:glauber}. This provides additional confidence 
that the observed entropy production rate is universal or scale-invariant. We used 
nonlinear optimization to determine the finite-size scaling exponents with the 
Nelder-Mead algorithm \cite{nelder}, employing all temperature ranges and entropy 
production data generated. The fitted exponents and parameters were identified 
for Glauber and Metropolis dynamics, respectively, summarized in Table \ref{tabfss}. 
This data collapse behavior supports the notion that entropy production is naturally 
bounded in driven nonequilibrium thermodynamic systems.

\section{Conclusions}

In this work, we introduce a thermal bath for the Ising–Conway 
Entropy Game (ICEg) with a novel coupling scheme that sustains 
nonequilibrium dynamics at a given temperature. As a pedagogically 
accessible zero-player game for statistical mechanics, playing the 
game under different settings provides valuable conceptual insights 
into entropy and its production rate in classical statistical physics, 
cooperative systems such as evolutionary games, and lattice 
gases—including Boltzmann lattice fluids. Our primary assumption 
is that the discrete game dynamics are representative of a physical 
system in thermal contact with a heat bath and evolve under self-driven 
dynamics, consistent with a fixed canonical ensemble.

The rate of entropy production increases more markedly with temperature 
for Glauber dynamics than for Metropolis dynamics, a statistically 
significant result. This suggests that Glauber dynamics may be more 
appropriate for lattice games with similar update rules.

In ICEg dynamics, computed entropy production reaches a maximum 
above a critical temperature, revealing a temperature-dependent 
dissipation limit that emerges from the interplay between local 
energy dissipation and kinetic constraints. This behavior, 
confirmed by comprehensive computational validation and supported 
by universal scaling collapse across multiple parameter regimes, 
suggests a fundamental constraint on entropy production in 
self-driven nonequilibrium systems—a principle that may 
extend beyond discrete lattice models to broader classes 
of stochastic dynamical systems. 

\section*{Acknowledgements}
We thank Y. S\"uzen for her kind support and encouragement.
We also thank the Scientific Python community \cite{vanrossum, numpy, scipy, matplotlib} 
for the powerful tools that enabled the development and numerical 
implementation of our results.

\section*{Declarations}
There is no conflict of interest. 
The dataset and source codes are available in the GitHub \cite{suzen25iceg} and 
Zenodo \cite{suzen25icegzen} repositories, ensuring full reproducibility 
and transparency.

\bibliographystyle{iopart-num}
\bibliography{suzen}

\end{document}